\documentclass[]{spie}

\usepackage[]{graphicx}
\usepackage{amsmath}
\usepackage{amssymb}
\usepackage[latin1]{inputenc}
\usepackage{floatflt}

\title{First Successful Adaptive Optics PSF Reconstruction at \\
W. M. Keck Observatory}

\author{Laurent Jolissaint\supit{a}, Chris Neyman\supit{b}, Julian Christou\supit{c}, Peter Wizinowich\supit{c} and Laurent Mugnier\supit{d}
\skiplinehalf
\supit{a}aquilAOptics, 1635 La Tour-de-Tr\^{e}me, Switzerland \\
\supit{b}W. M. Keck Observatory, Hilo, Hawai'i, USA\\
\supit{c}Gemini Observatory, Hilo, Hawai'i, USA\\
\supit{d}ONERA/DOTA, Ch\^{a}tillon Cedex, France}

\authorinfo{Further author information: e-mail: laurent.jolissaint@aquilaoptics.com, web page: www.aquilaoptics.com}

\begin{document} 
\maketitle 

\begin{abstract}
We present the last results of our PSF reconstruction (PSF-R) project for the Keck-II and Gemini-North AO systems in natural guide star mode. Our initial tests have shown that the most critical aspects of PSF-R are the determination of the system static aberrations and the optical turbulence parameters, and we have set up a specific observation campaign on the two systems to explore this. We demonstrate that deformable mirror based seeing monitor works well, and 10\% accuracy is easily obtained. Phase diversity has been demonstrated to work on sky sources. Besides, residual phase stationarity is an important assumption in PSF-R, and we demonstrate here that it is basically true. As a result of these tests and verifications, we have been able for the first time to obtain a very good PSF reconstruction for the Keck-II system, in bright natural guide star mode.\\
\textbf{This paper is to be published in the proceedings of the 2nd AO4ELT conference held in Victoria, BC, Canada, September 26-30, 2011}.
\end{abstract}


\section{Introduction}\label{sec:1}

Reducing adaptive optics (AO) data often requires the knowledge of the point spread function (PSF) associated to the AO run. Due to the sensitivity of the AO system performance to the optical turbulence conditions (seeing), the fast variation of these conditions, and the complexity of the PSF structure, it is not possible to build generalist PSF models that would well represent the current PSF, as it is the case in seeing limited mode. Getting the PSF from a point source image recorded before/after the AO run would be a natural option, but the seeing changes too rapidly and pre/post-run PSF are of limited interest. An ideal situation occurs when a bright, isolated star image is available close to the science object of interest, but this case is rare in practice. A general method, independent of the field and the seeing conditions, is therefore required.
 
V\'{e}ran et al. \cite{veran97} have proposed and demonstrated successfully a PSF reconstruction (PSF-R) technique based on the AO loop data. The concept is simple: first, the WFS measures the low order\footnote{throughout this paper, low order modes indicates the modes (Zernike, Influence Functions etc.) that can be sensed and corrected by the system, and high order modes are all the others - in particular the turbulent aberrations at spatial frequencies above the system cutoff frequency 0.5/pitch.} residual wavefront, so it must be possible to evaluate from these measurements the contribution of the wavefront residual to the long exposure PSF; second, as the deformable mirror (DM) shape is set to compensate for the incoming turbulent wavefront, it must be possible as well to determine, by looking at the DM commands statistics, the seeing parameters associated to the AO run, and, in turn, evaluate the amplitude of the high order aberrations that went through the system, which are, by nature, uncorrected. This idea was developed by V\'{e}ran et al. for a curvature sensing (CS) WFS system and tested on PUEO, a CS system installed on the CFHT telescope. Later, we adapted V\'{e}ran's method to the Shack-Hartmann type WFS \cite{jolissaint04} and applied it on ALTAIR, the Gemini-North AO system, and recently on the Keck-II telescope AO system \cite{jolissaint10}. The PSF-R theory will not be recalled here, the interested reader is invited to consult references \cite{veran97} and \cite{jolissaint04}.

Other options might be possible (but have not been tested yet): for instance, a parametrized PSF model (look-up table) that would be build from sky measurements, where the parameters would be the seeing and the most influential loop parameters (gain, frame rate); an other, albeit more expensive option would be to design a dedicated PSF measurement device within the instrument optics, that would pick up part of the guide star(s) light, purposely propagated beyond the AO bench. This might be a back-up solution if PSF-R algorithms for multi-(laser) guide stars, multi-DM systems are proven too difficult to devise or implement.

In any case, our group has chosen to proceed with V\'{e}ran's method, and finalize its implementation on the Gemini-N (Altair) and Keck-II AO systems. Our first experience \cite{jolissaint10} has demonstrated that the most critical aspects of PSF-R are the determination of the system's static aberrations and the optical turbulence parameters (seeing angle $\omega_0$, outer scale $L_0$) as seen from the telescope. The PSF structure is very sensitive to these parameters, and the difficulty we faced with determining these parameters with accuracy prevented us to reconstruct the PSF on these systems, until now. We have therefore concentrated our efforts on these two aspects, and we report our results here.
\begin{floatingfigure}[p]{65mm}
\centering
\includegraphics[width=60mm]{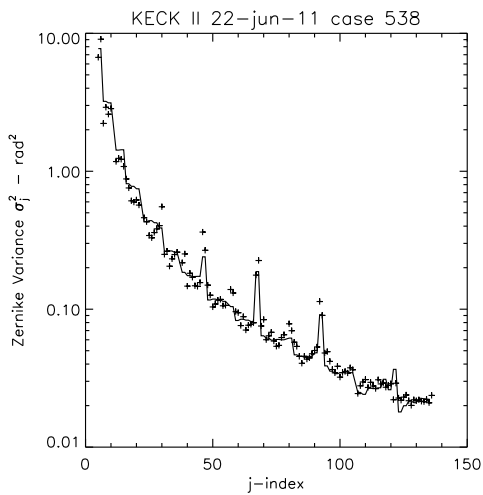}\hspace{2mm}
\caption{Variance of the Zernike coefficients from j-indexes 5 to 136, for one of the closed loop cases of the 22-June-2011 night. Purely radial modes (m=0) are not shown nor used. Continuous line: Noll's model, scaled ($D/r_0=55.2$) to the DM commands variances, projected in the Zernike space, and adjusted for the outer scale (17 m here). Central obscuration ($\epsilon=0.304$) is taken into account in the projection, and is responsible for the jumps at some j-indexes.}
\label{fig:01}
\end{floatingfigure}

For the seeing estimation, we use the approach proposed by V\'{e}ran et al.: extracting the average Fried parameter $r_0$ and the optical turbulence outer scale $L_0$ from the DM commands statistics. The advantage is that this method is telescope bound: getting the seeing from a monitor not collocated with the telescope optical axis (as is the case with the Mauna Kea summit DIMM/MASS seeing monitor) does not deliver any useful seeing information for the PSF-R procedure, as we will see. The question, though, was how accurate can be the DM-based seeing monitor ? We have set up an experiment at Keck-II to answer this question, and our method and results are reported in section \ref{sec:2}. Our work complements an initial study made by Schoek et al. \cite{schock03} who used open loop WFS measurements to retrieve the seeing parameters (but of a limited interest in PSF-R because simultaneity with the closed loop session is required).

Regarding the static aberrations determination, we made use of a phase diversity (PD) approach developed by Mugnier et al. \cite{mugnier08}, specifically adapted to the use of AO corrected sky source images. We have indeed realized that for reasons not understood yet, the internal static aberrations compensation procedure implemented in the two systems, both using internal calibration sources, does not correct all the aberrations seen at the detector level. In order to capture the aberrations of the whole optical train, from the telescope pupil to the detector plane, we required a light source above the telescope, i.e a sky source. Our on-sky PD results are presented in the section \ref{sec:3}.

Finally, there is the question of the residual phase stationarity approximation: this is very central in V\'{e}ran's approach, and allows to write the optical transfer function (OTF) of the whole optical system as the product of the telescope+instrument OTF and an AO-OTF filter, $\mathrm{OTF}_{\mathrm{all}}=\mathrm{OTF}_{\mathrm{tel}}\,\mathrm{OTF}_{\mathrm{AO}}$. Telemetry data is used to compute the AO-OTF, while the telescope (or static) OTF needs to be measured by other means (phase diversity for instance). It can be demonstrated that this approximation generates an underestimation of the global OTF, so the reconstructed PSF is pessimistic with respect to the real PSF. We have the mean to test this assumption by examining the statistics of the reconstructed residual wavefront within the pupil. Our analysis (section \ref{sec:4}) shows that stationarity is basically verified, so the approximation is valid.

\section{Testing the DM seeing monitor with open loop seeing measurements in the focal plane}\label{sec:2}

If the DM commands are expressed in equivalent Zernike coefficients, the variance of these Zernike coefficients should closely follow the distribution given by Noll \cite{noll76} (correcting for the outer scale \cite{jolissaint04}). Our measurements at Keck-II demonstrate that it is indeed the case (see Fig. \ref{fig:01}). This being said, there are some practical difficulties in this DM-based seeing monitor that can impact the accuracy: (1) contribution of non-turbulent dynamical aberrations to the DM commands, due to the wind (vibrations, jitter, primary mirror warping etc.), focus tracking errors, etc.; (2) contribution of the WFS noise to the DM commands; (3) likelihood of the outer scale of optical turbulence model and its damping effect on the Zernike variances. Knowing that the seeing is a critical parameter, we needed to verify the DM-based seeing values with an independent measurement of the seeing seen at the telescope focus, at the same time.

We selected a bright natural guide star (NGS) to allow for a low (negligible) noise level, then alternated, as quickly as possible, 30 seconds closed-loop (CL) and open-loop (OL) sessions. During the CL sessions, we recorded the DM telemetry, from which we extracted $r_0$, and followed the procedure described in \cite{jolissaint04} to determine $L_0$. During the OL sessions, we recorded the seeing limited PSF with the NIRC2 camera (Near Infrared Camera 2) - note that at Keck-II, OL actually means tilt correction: we took this into account in our seeing values comparisons. The full CL/OL sessions lasted for about an hour, on June 22 \& 23, 2011, and a short 15 minutes test on October 13, 2011. Details of the reduction and processing of the OL/CL data will be described in a forthcoming paper.
\begin{floatingfigure}[p]{60mm}
\centering
\includegraphics[width=57mm]{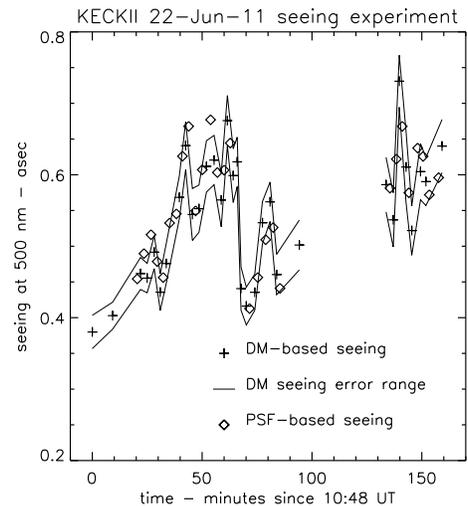}\hspace{2mm}
\caption{Seeing monitor campaign results - June 22, 2011, Keck-II telescope. DM-based seeing time evolution as compared to the open loop seeing. Continuous line indicates the 1-$\sigma$ error.}
\label{fig:02}
\end{floatingfigure}
The reduced DM-seeing and NIRC2-seeing time series are compared in Fig. \ref{fig:02}. The agreement is remarkable (the linear correlation coefficient between both series is $r = 0.95$). Looking at the statistics of the difference between the DM and NIRC2 values, over the 146 measurements we made for the June and October nights, we find that the relative accuracy of the DM-seeing monitor is better than 5\% 2/3rd of the time, and better than 10\% 92\% of the time. The maximum relative error we had on these 146 sets was 16\%. We also compared our DM-based seeing with the values given by the MASS/DIMM seeing monitor\footnote{http://wxws.ifa.hawaii.edu/current/seeing/}, at the same time, installed at the summit, 580 meters from the Keck observatory. We find that the linear correlation is rather poor, in the range $r = 0.4-0.5$.

We had no independent measurement of the outer scale, so it was not possible to assess the accuracy of its determination. This being said, according to A. Ziad \cite{schock01}, our values - median $L_0$=22 m, $\sigma_{L_0}$=6 m, range 10-45 m - are clearly compatible with the results of a GSM (Generalized Seeing Monitor) optical turbulence characterization campaign at Mauna Kea in Nov/Dec 2001 (median $L_0$=18 m, 2/3rd of the values within a 7-50 m range). Our conclusion is that the closed loop DM-based seeing monitor technique works well, for estimating both the seeing value $\omega_0$ and the optical outer scale of turbulence, $L_0$. There is therefore no need to implement independent seeing monitors at telescopes equipped with AO systems: the DM telemetry can be used for this.

\section{Retrieving the AO system and instrument static aberrations from on-sky phase diversity}\label{sec:3}

\subsection{Introduction}\label{sec:3.1}

An AO system is only capable of correcting the aberrations that can be seen by the WFS. Post-AO aberrations cannot be compensated, neither can the aberrations occurring only in the WFS arm, after the beamsplitter (the so-called non common path aberrations, NCPA). Pre-AO static aberrations at spatial frequencies above the WFS spatial sampling frequency 0.5/pitch are not seen either. NCPA aberrations, though, can be pre-compensated by a constant offset applied to the DM commands. This requires a measurement of these static aberrations, on both the instrument and the WFS arm . Procedures are implemented in most AO systems to measure these NCPA, using internal calibration light sources. This being said, these NCPA compensation schemes, sometime, do not work as well as expected, and there can be a some level of low order static aberration in the focal plane.

This is basically the issue we are currently facing at the Keck-II AO system, where according to our initial analysis \cite{jolissaint11}, a low order aberration in the range 100-300 nm seems to be missing in the reconstructed PSF w.r.t. the actual sky PSF. There is an image sharpening procedure implemented on the Keck-II system, which makes use of an internal pre-AO calibration source. This procedure is based on a PD algorithm\footnote{the basic concept is simple: it can be shown that one can extract the common aberration from two PSF which differs only from an additional and known aberration. In general one of the PSF is on-focus, the other is defocused, from a known amount.}, the outcome of which is a phase map that is loaded on the DM as a permanent offset. Unfortunately, for reasons not clear yet, this pre-compensation system does not produce the expected image quality in sky observing mode. For this reason, we have decided to short-cut this image sharpening procedure and implement our own PD procedure, on a bright sky source, in closed loop mode, in order to capture all the aberrations from the entrance pupil down to the focal plane. The method and code we are using were developed by Mugnier et al. \cite{mugnier08}. This algorithm has only been demonstrated on simulated images, so this is the first time - as far as we know - that PD is applied on closed loop sky data.

\subsection{Implementing phase diversity: lessons learned}\label{sec:3.2}

In principle, two PSF with a phase difference of about $2\pi$ peak-to-valley are enough for PD. We nevertheless took a series of gradually increased defocused PSF to visually explore the PSF change. We took sets of seven PSF, three on each side of the best focus plane. We have immediately faced a certain number of difficulties:

\subsubsection*{Seeing variation}

The seeing cannot be expected to be stable from a defocused PSF acquisition to another, the delay generally being on the order of 2 to 5 minutes (exposure, detector reading, focus change). For instance, for one of our defocused PSF sets, the seeing varied in the range 0.7" to 1". Now, with the PD approach, it is assumed that the difference between the defocused PSF is only generated by the defocus, and that the imaged object remains the same. In our case, the imaged object is the seeing limited PSF, corrected for the low order aberrations. What is left in this PSF are the seeing wings, plus some low order aberrations residuals. The amplitude of the wings is proportional\footnote{in a first approximation, the PSF shape - left aside the diffractions rings - is proportional to the residual phase spatial power spectrum} to the factor $r_0^{-5/3}$, and the width proportional to $r_0$. Therefore, with the natural variation of the seeing, the imaged object cannot be considered identical from PSF to PSF.

The consequence is that the PD algorithm will misinterpret the seeing-based PSF variation as coming from a non existent static aberration, with a radial symmetry because the long exposure AO PSF is mostly axisymmetric. This is exactly what we experienced: purely radial modes were dominant in the distribution of the Zernike modes when all the defocused images were used. We decided therefore to use our DM-based seeing monitor and compute the seeing angle for all the seven defocused images, and keep only PSF pairs for which the relative seeing difference was not larger than 10\%, and the phase defocus was close to $2\pi$. Doing this basically brought back all the $m=0$ Zernike coefficients to about the same values than the other modes, except for the defocus term $a_4$ which was still very significant.

In any case, this exercise just demonstrated that it is crucial, when doing on-sky PD, to have at one's disposable a reliable (say better than 10\% accuracy) estimate of the seeing associated with the PSF, and keep only the PSF with similar seeing angle. If PD is very critical for the system calibration, one option would be to implement a dedicated beamsplitter design in front of the camera to record the on-focus and defocused PSF simultaneously. Otherwise, lots of PSF must be recorded at each defocus distance in order to be certain that there will be matching values of the seeing for at least two PSF.

\subsubsection*{Optical system design parameters}

The second difficulty is that the setting of the PD algorithm requires the knowledge of the geometry of the optical beam, with high accuracy: f-number in the image plane, pupil central obscuration, pupil boundary shape, and the exact amount of defocus that was introduced from a PSF to another. Unfortunately, these numbers are not necessarily known with certainty, because the AO or instrument documentation does not always give these numbers, or they cannot be find easily or with accuracy. Sometimes, the camera is mechanically fixed and cannot be shifted along the optical axis to emulate a defocus, and in this case the defocus has to be introduced either by applying a constant $Z_4$ term to the DM, or by defocusing the WFS, automatically introducing an opposite $Z_4$ term in the corrected beam.

Again, in these two cases, the relationship between the f-number and the actual defocus introduced at the focal plane depends on the system geometry and lack of proper or accurate documentation is possible here too. Introducing the wrong geometry to the PD algorithm generates spurious purely radial (m=0) aberrations in the solution. This is what we have experienced (excessive m=0 coefficients) at Gemini-N ... until we found that we needed to introduce a 0.5 correction factor in the f-number of the Gemini-N optics. So, the second lesson with PD is that the optical system geometry must be well known.

\subsubsection*{Phase diversity code parameters}

Finally, the PD algorithm itself has its own limits. Getting the phase from a set of two PSFs is an inverse problem, which is solved by the minimization of an error metric. In our code this metric can be interpreted as deriving from a Joint Maximum A Posteriori approach \cite{blanc03}. While noise is taken into account in the data model and thus controlled in the inversion, modeling errors can have a strongly detrimental effect on the estimation. Such errors are in particular due to an imperfect calibration of the system, e.g. of the pupil size and shape, of the defocus distance, of the camera pixel scale, as studied in \cite{blanc03b}.

One could contemplate regularizing the phase estimate using prior knowledge on the spatial frequency content of the aberrations to be estimated. In the absence of such prior knowledge, we have to limit the number of modes we want to reconstruct. This limit results from a compromise between phase modeling errors (if the number of modes is too small) and noise amplification (if it is too large). From experience, we chose to reconstruct 6 radial orders -- and we found that reconstructing more modes was not making the fit of the solution closer to the measurements. We plan to study the noise propagation \cite{meynadier99} more finely in the future for this Keck data.

\subsection{Cross checking our phase diversity code with another and equivalent code}\label{sec:3.3}

F. Rigaut shared his PD tool (OPRA) with us, a code written in the \texttt{yorick} language, in order to do cross-checks with Mugnier's tool. On calibration sources, at Gemini-N, both codes produced almost exactly the same results, the linear correlation coefficient between both solutions being an excellent $r=0.985$. So, there is no doubt that these codes do produce realistic solutions, provided that the sensitive issues discussed above are taken into consideration.

\subsection{Phase diversity at Gemini-North: internal calibration source results}\label{sec:3.4}

Our first PSF-R tentative for Altair at Gemini-N\cite{jolissaint11} have clearly shown that there is a significant amount of unexpected residual WFE in the whole system. We have identified that the major offenders are post-AO vibrations, and a high order, pre-AO telescope aberration generated by the M2 mirror support structure. Besides, PD experience with other AO systems (NACO\cite{hartung03}) demonstrates that filters are a potential source of aberrations. In order to evaluate this for Altair, we recorded a set of 15 focused/defocused PSF on the Near Infrared Imager (NIRI) (K-prime filter g0206, $\lambda=2120$ nm), using the calibration lamp as the point source. The M2 aberration therefore had no influence here. Exposure time was long enough to consider the vibrations to produce simply a smoothing of the PSF. Our conclusions:
\begin{itemize}
\item there is a Zernike defocus error of $a_4=62$ nm, which means that the real NIRI focus is 0.426 mm away from the notional focus. Such a defocus is not outrageous and is easy to correct.
\item the overall WFE (155 nm w/o the defocus) is dominated by astigmatism (113 nm - very apparent in the defocused images), then coma (57 nm), then spherical (29 nm) - these aberrations make for 85\% of the total WFE (in quadrature, w/o defocus) - the other aberrations are on the order of $\pm10$ nm. These results are very similar, in modes and amplitude, to what was measured on the NACO system \cite{hartung03}, indicating, possibly, a similar source for this post-AO aberration: the filters.
\end{itemize}
In any case, the Strehl associated with this reconstructed phase is 81\% in K-band, and this is clearly insufficient to match for the missing Strehl between the reconstructed AO PSF and the detector PSF (which is half the reconstructed Strehl). In other words, post-AO static aberrations cannot b a dominant source of static WFE. So we are clearly left with the vibrations and the M2 issue as the main aberration source.

\subsection{Phase diversity at Keck-II: on-sky results}\label{sec:3.5}

With these good PD results at Gemini-N, we were confident to try the PD technique on sky data, at Keck-II. We knew from our initial PSF-R experiment that something was not right with the image sharpening procedure, in a sense that when the calibrated system is turned to the sky, there is still (but not always) a significant amount of low order static error. So we took two sets of seven defocused PSF, for the filters FeII (1646 nm) and $[\gamma]$ (2169 nm), in closed loop, keeping the AO telemetry for each PSF. Also, in order to minimize the source of potential pupil definition errors, we decided to make the Keck-II pupil circular, by using the NIRC2 dedicated 8.72-m circular pupil mask, with a 3-m central obscuration. A cross-check of the two codes (Rigaut's and Mugnier's) on Keck-II calibration source PSF produced again very similar results.

We have selected PSF couples for which the focus difference was large and the seeing angle was as close as possible (less than 10\% difference). Six radial orders were sufficient to reconstruct the static phase. We found an overall WFE of 195 nm in $[\gamma]$ and 208 nm in Fe-II, \textit{strongly dominated by a defocus term}: -146 nm in $[\gamma]$ and -183 nm in FeII ! The other terms are the spherical $Z_{11}$ at 30 nm followed by $Z_{22}$ at -70 nm. The 195-208 nm WFE matches well the missing WFE we expected from our previous PSF-R experiment. So it seems that in the Keck-II case, we are now able to see most of the static aberration.

\section{Validity of the residual phase stationarity assumption}\label{sec:4}

\mbox{}
\begin{floatingfigure}[p]{75mm}
\centering
   \includegraphics[width=55mm]{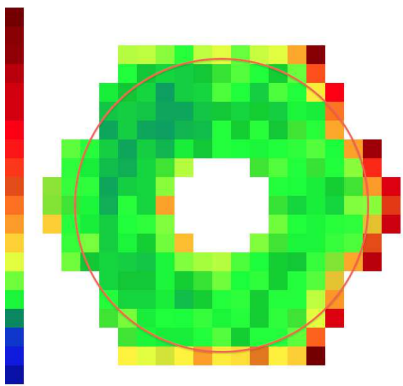} 
\caption{Pupil map of the variances of the DM commands residuals (WFS slopes were transformed into DM commands). Each pixel represents one actuator. Color code is indicated, white corresponds to 0, dark red to the maximum. The circle shows the boundary of the 8.72-m pupil mask.}
\label{fig:04}
\end{floatingfigure}
In order to evaluate the corrected phase homogeneity across the pupil, we have build pupil maps representing the variance of the residual DM commands at each actuators (June 22, 2011 nights).
Considering the actuators that are within the hexagonal pupil, i.e. the ones that have an impact on the image formation, the residual phase is essentially stationary everywhere (83\% of the pupil surface), except on a few actuators on the edge - Fig. \ref{fig:04}. Inside the 8.72-m mask, the residual phase is stationary everywhere (red circle). The stationary assumption is, basically, valid.

We can therefore safely proceed with the separation of the global OTF into a telescope and AO-OTF. We expect that the higher the actuators density, the better the stationary assumption, because for systems with a large number of actuators, the low variance actuators will largely dominate. In other words, reconstructing the PSF for AO systems with a high actuators density (ExAO systems, AO on extremely large telescopes) can also make use of the OTF separation paradigm.


\section{PSF-R at Keck-II in Natural Guide Star Mode}\label{sec:5}

The way is paved now for a successful PSF reconstruction on the Keck-II system, in NGS mode: we have a reliable local seeing monitor to accurately reconstruct the fitting error and the WFS aliasing components of the AO-OTF; a reliable way to determine the static aberrations of the whole system, so we can build a model of the telescope+instrument OTF; and we know that the fundamental assumption of the PSF-R method - phase stationarity - holds true, and we have a method to compute the $U_{i,f}$ functions\footnote{we exploit influence functions symmetries, and the fact that pupil edges can be neglected in the stationary phase approximation}.

\begin{figure}[t]
   \centering
   \includegraphics[width=58mm]{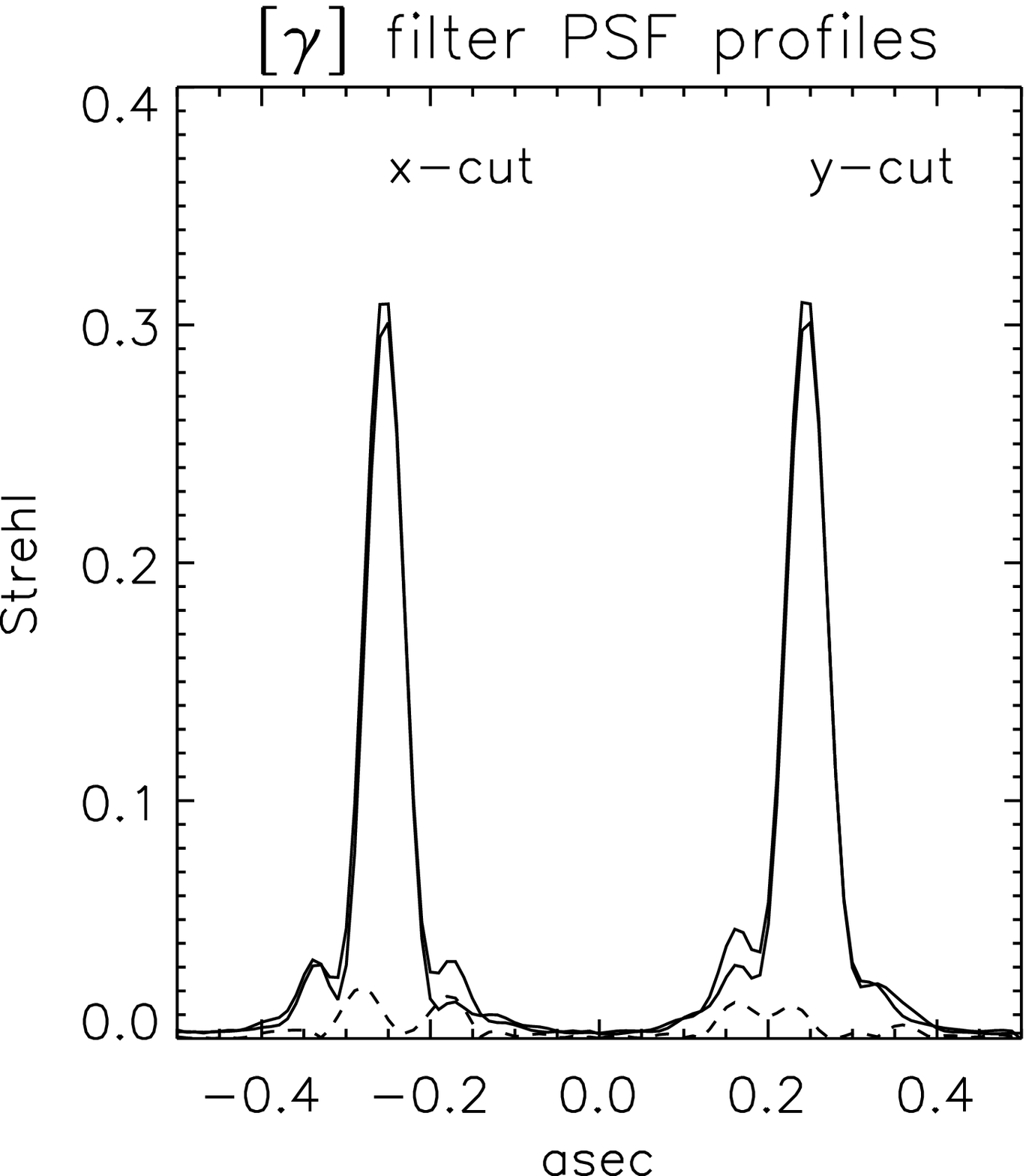}\hspace{5mm}
   \includegraphics[width=58mm]{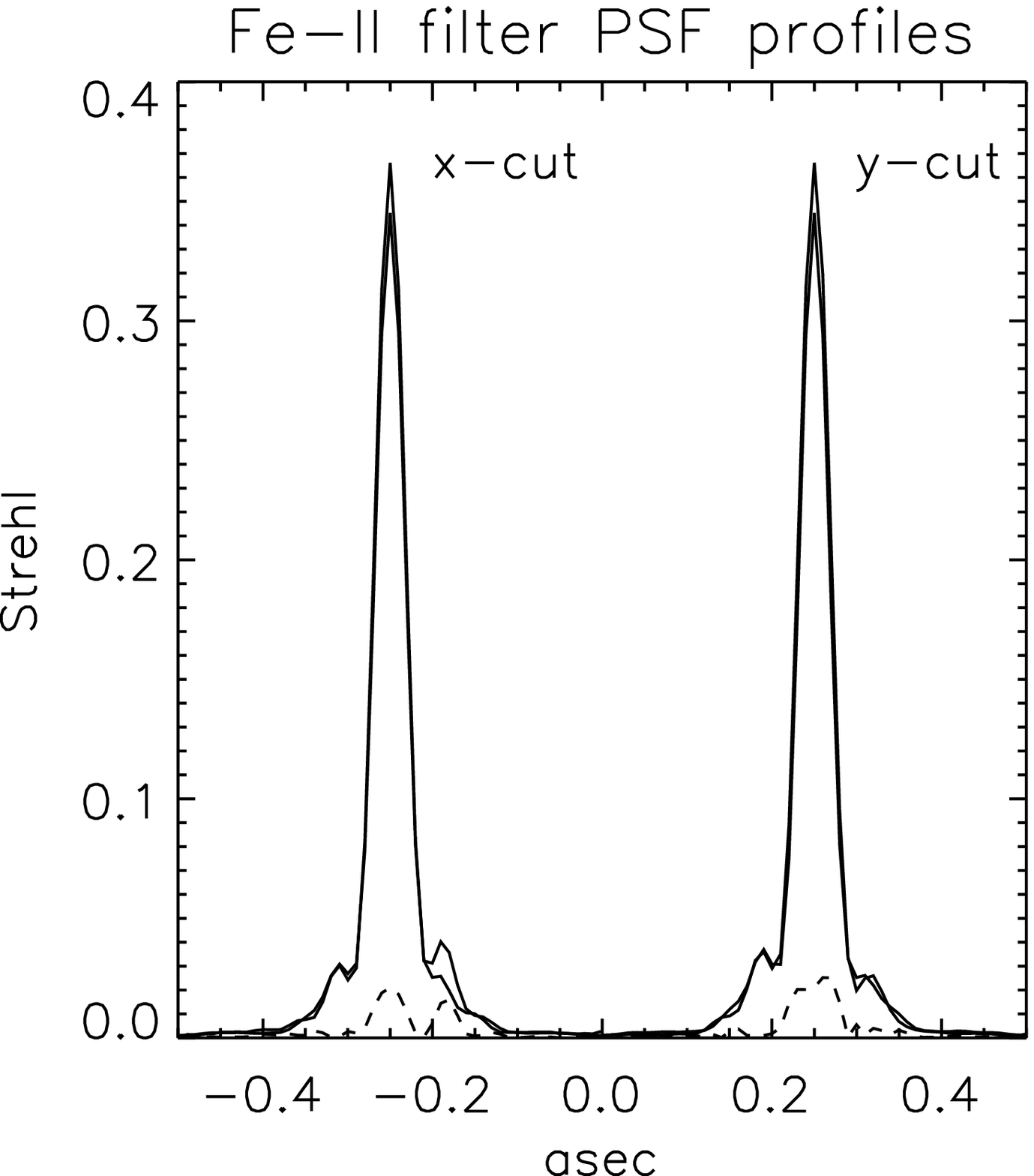}
   \caption{Horizontal and vertical cut of the sky and reconstructed PSF, 23-Jun-2011 data, $[\gamma]$ and Fe-II filters. Keck-II AO system, bright NGS mode. Dashed lines indicate the absolute value of the PSF profile differences. FWHM and Strehl are given in table \ref{tab:01}.}
   \label{fig:psfprofiles}
\end{figure}
\mbox{}
\begin{floatingtable}{
\centering
\begin{tabular}{rlccc}
			& fwhm			& $\alpha$ 	& SR 	& $w_0$ \\
			& masec		& -	& [\%]	& ["]    \\ \hline
$[\gamma]$ s& $53/56$ 	& $75^{\circ}$	& 31 	& 1.05   \\
$[\gamma]$ r	& $51/58$ 	& $62^{\circ}$	& 30 	& --     \\ \hline
FeII s		& $41/42$	& $86^{\circ}$	& 37 	& 0.76   \\
FeII r		& $41/41$	& $0	^{\circ}$	& 35 	& --
\end{tabular}}
\caption{FWHM min/max/angle and Strehl measured on the sky (s) and reconstructed (r) PSF, for both the $[\gamma]$ (2169 nm) and Fe-II (1646 nm) filters. DM seeing is indicated. 23-Jun-2011 PSF-R experiment, Keck-II.}
\label{tab:01}
\end{floatingtable}
We had time for only three PD + PSF-R data acquisition (defocused PSF and loop telmetry): 1st data set is with the $[\gamma]$ filter and the 8.72-m pupil mask set, 2nd data set is with the Fe-II filter and the 8.72-m pupil mask, 3rd data set is with the same filter, and the full hexagonal pupil (no mask). PD was not successful on the later case (essentially for reasons indicated in section \ref{sec:3.2}). We selected bright NGS targets, to minimize the impact of the WFS noise. Static phase was retrieved for the two circular mask cases (numbers, see section \ref{sec:3.5}), and the AO-OTF was reconstructed from the telemetry for the central (focused) PSF.
The following sources of residual aberrations were included into our OTF model (basically, all sources that we were able to identify and model at this time): (1) system's static aberrations, from PD - see section \ref{sec:3.5}; (2) high order optical turbulence aberrations, or fitting error, and WFS aliasing, from the $r_0$ we got from the DM-seeing monitor; (3) low order wavefront residuals (from the WFS telemetry) - WFS noise was neglected; (4) residual jitter (from the WFS telemetry); and the detector modulation transfer function (MTF).
The Strehl \& FWHM of the sky and reconstructed PSF are given in table \ref{tab:01}. The agreement is excellent: the Strehl error is on the order of 5\% for both filters, and the FWHM error is basically un-significant. PSF profile comparison are shown in Fig. \ref{fig:psfprofiles}: the detailed PSF structure shows some differences, and this is certainly due to the uncertainties in the determination of the static aberrations. Indeed, the residual turbulence part of the corrected PSF is in principle axisymmetric, therefore the un-even PSF difference we see is necessarily coming from the static aberration. Surprisingly, the performance in the Fe-II filter was better than in $[\gamma]$: this demonstrates the impact of the seeing on the residual phase - the seeing was far stronger during the $[\gamma]$ AO run than during the Fe-II run (table \ref{tab:01}, right column).

Table \ref{tab:02} (next page) gives the distribution of the overall WFE (from the sky Strehl using the Mar\'{e}chal law) in the different error posts. The static aberration dominates in the Fe-II filter, and the residual turbulence WFE dominates in $[\gamma]$. This is a strong indication that our PSF-R procedure does work: indeed, if the AO-OTF model was wrong, then we would not be able to reproduce with such accuracy the Strehl, unless an error on the static aberrations was compensating for an error in the AO-OTF, but this is unlikely.

\section{Conclusions}\label{sec:6}

\begin{floatingtable}[r]{
\centering
\begin{tabular}{lrr}
error [nm]	& $[\gamma]$	& Fe-II \\\hline
static		& 195		& 208 \\
fitting		& 167		& 127 \\
aliasing		&  91		&  73 \\
low order	& 263		&  91 \\\hline
total		& 379		& 270 \\\hline
\end{tabular}}
\caption{Distribution of the overall WFE on the main error posts - June 23, 2011 PSF-R Keck-II experiment. Low order WFE includes both the jitter and the other low order terms.}
\label{tab:02}
\end{floatingtable}
We claim a few premieres in this paper: demonstration of the validity of the DM-based seeing monitor in closed loop, first on-sky demonstration of sky-based phase diversity, demonstration of the validity of the stationary phase assumption, and finally, first successful PSF reconstruction on the Keck-II telescope. Static aberrations determination is clearly an aspect of the PSF-R procedure as critical as the residual turbulence AO-OTF determination. Our phase diversity procedure, while giving encouraging preliminary results, is still relatively "fragile" and some work is needed to make it work for the Keck-II hexagonal pupil. Besides, we have specifically neglected the impact of the WFS noise in our study, by selecting bright NGS. In the next steps of this project, we will gather dimmer NGS data, and apply the noise filtering procedure described in \cite{jolissaint04}.
\vspace{2mm}

\acknowledgments
 
L.J. would like to thank W. M. Keck Observatory, Hawaii (United States) and Gemini Observatory for their generous support during this study. The authors would like to thank also Fran\c{c}ois Rigaut and Benoit Neichel for many fruitful discussions concerning phase diversity procedures at Gemini Observatory.

The Gemini Observatory is operated by the Association of Universities for Research in Astronomy, Inc., under a cooperative agreement with the NSF on behalf of the Gemini partnership: the National Science Foundation (United States), the Science and Technology Facilities Council (United Kingdom), the National Research Council (Canada), CONICYT (Chile), the Australian  Research Council (Australia), Minist\'{e}rio da Ci\^{e}ncia e Tecnologia (Brazil), and Ministerio de Ciencia, Tecnologia e Innovaci\'{o}n Productiva (Argentina).

Data presented herein were obtained at the W. M. Keck Observatory, which is operated as a scientific partnership among the California Institute of Technology, the University of California, and the National Aeronautics and Space Administration. The observatory was made possible by the generous financial support of the W. M. Keck Foundation. The authors recognize and acknowledge the significant cultural role and reverence that the summit of Mauna Kea has always had within the indigenous Hawaiian community. We are most fortunate to have the opportunity to conduct observations from this mountain.


\begin{thebibliography}{}
	\bibitem{veran97} J-P. V\'{e}ran, F. Rigaut, H. Maitre, D. Rouan, JOSA A, \textbf{11}, (1997), 3057-3069
	\bibitem{jolissaint04} L. Jolissaint, J-P. V\'{e}ran, J. Marino, SPIE Proceedings, \textbf{5490}, (2004), 151-163
	\bibitem{jolissaint10} L. Jolissaint, J. C. Christou, P. Wizinowich, E. Tolstoy, SPIE Proceedings, \textbf{7736}, (2010), 77361F-77361F-12
	\bibitem{schock03} M. Schock, D. Le Mignant, G. Chanan, P. Wizinowich, and M. van Dam, ApOpt {\bf 42}, (2003), 3705
	\bibitem{mugnier08} L. Mugnier, J.-F. Sauvage, T. Fusco, A. Cornia and S. Dandy, Optics Express, {\bf 16}, (2008), 18406-18416
	\bibitem{noll76} Noll, R. J., JOSA, {\bf 66}, (1976), 207-211
	\bibitem{schock01} M. Schock, A. Ziad, M. Chun, G. Chanan, A. Lewis, A. Hatahiyama, R. Wainscoat, CELT Project, Internal Publication, (2001)
	\bibitem{jolissaint11} L. Jolissaint, J. C. Christou, C. Neyman, in Adaptive Optics: Methods, Analysis and Applications, OSA Technical Digest (CD), (2011), paper APDP2
	\bibitem{blanc03} A. Blanc, T. Fusco, M. Hartung, L. Mugnier, and G. Rousset, A\&A, {\bf 399}, (2003), 373-383
	\bibitem{blanc03b} A. Blanc, T. Fusco, M. Hartung, L. M. Mugnier, and G. Rousset, A\&A {\bf 399}, (2003), 373-383
	\bibitem{meynadier99} L. Meynadier, V. Michau, M.-T. Velluet, J.-M. Conan, L. M. Mugnier, and G. Rousset, Appl. Opt. {\bf 38}, (1999), 4967-4979
	\bibitem{hartung03} M. Hartung, A. Blanc, T. Fusco, F. Lacombe, L. Mugnier, G. Rousset and R. Lenzen, A\&A, {\bf 399}, (2003), 385-394
\end{thebibliography}
\end{document}